\documentclass[%
 reprint,
 amsmath,amssymb,
 aps,
]{revtex4-1}

\usepackage{graphicx}
\usepackage{dcolumn}
\usepackage{bm}
\usepackage{float}
\usepackage{amsmath}
\usepackage{epstopdf}

\begin{document}

\preprint{AIP/123-QED}

\title[]{Radio-over-fiber using an optical antenna based on Rydberg states of atoms}

\author{A. B. Deb}
 \email{amita.deb@otago.ac.nz}
\author{N. Kj{\ae}rgaard}%

\affiliation{ QSO - Centre for Quantum Science, Dodd-Walls Centre and Department of Physics, University of Otago, Dunedin 9010, New Zealand.
}%

\date{\today}

\begin{abstract}

We provide an experimental demonstration of a direct fiber-optic link for RF transmission (``radio-over-fiber") using a sensitive optical antenna based on a rubidium vapor cell. The scheme relies on measuring the transmission of laser light at an electromagnetically-induced transparency resonance that involves highly-excited Rydberg states. By dressing pairs of Rydberg states using microwave fields that act as local oscillators, we encoded RF signals in the optical frequency domain. The light carrying the information is linked via a virtually lossless optical fiber to a photodetector where the signal is retrieved. We demonstrate a signal bandwidth in excess of 1 MHz limited by the available coupling laser power and atomic optical density. Our sensitive, non-metallic and readily scalable optical antenna for microwaves allows extremely low-levels of optical power ($\sim 1\, \mu$W) throughput in the fiber-optic link. It offers a promising future platform for emerging wireless network infrastructures.

\end{abstract}

\pacs{Valid PACS appear here}
\maketitle

The integration of fiber-optical and wireless networks has underpinned the modern information age. In recent years, the proliferation of smart communication devices and appliances have fuelled an unprecedented demand for high bandwidth wireless networks. The emerging fifth-generation (5G) mobile network, for instance, will feature the 20 GHz - 30 GHz milli-meter wave band \cite{Bohata2018,Kanno2018, Kanno2011}. High-frequency millimeter waves, however, are prone to significant attenuation and deflection by obstacles, including air and rain \cite{Sizun2006}. Transmission span can be extended by combining free-space propagation with guided propagation in e.g., coaxial cables. Metallic coaxial cables are, however, extremely lossy - typically $>1\,$dB/meter at 25\,GHz - requiring amplification stages every few tens of meters. A highly prospective solution to this is ``radio-over-fiber" (RoF) \cite{Bohata2018,Kanno2018,Novak2016,Kuboki2018,Tien2015,Wake2010}, where RF signals are encoded in laser light and transported via fiber-optic channel(s). Optical fibers offer extremely low levels of loss - typically 0.3~dB/kilometer requiring no amplification stages over hundreds of kilometers. They also provide exquisite electrical isolation and are very lightweight.

Conventional RoF architectures comprise a signal source (Figure 1a), an rf-to-optical encoder, a fiber-optic link, a photodetection stage and demodulation electronics \cite{note} (Figure 1b). RF-to-optical encoding is done either by intensity modulation of a laser source via e.g. controlling laser current \cite{Bohata2018} or intensity/phase modulation of light using electro-optical crystals \cite{Kanno2011}, both using guided RF signals. This requires direct electrical connection and complex amplification and filtering stages. Efficient conversion of microwave signals into an optical signal is an outstanding challenge due to small Kerr nonlinearities of conventional crystals, often necessitating high optical pump power \cite{Davis1996}. This renders high modulation-depth and efficient optical encoding of RF signals a difficult task. A sensitive optical receiver antenna that enables direct encoding of \emph{free-space} RF signals eliminates the need for any electrical contacts at the receiver end, enabling a far simpler and more versatile RoF architecture as shown in Figure 1c. 


\begin{figure*}
\includegraphics[width=165mm]{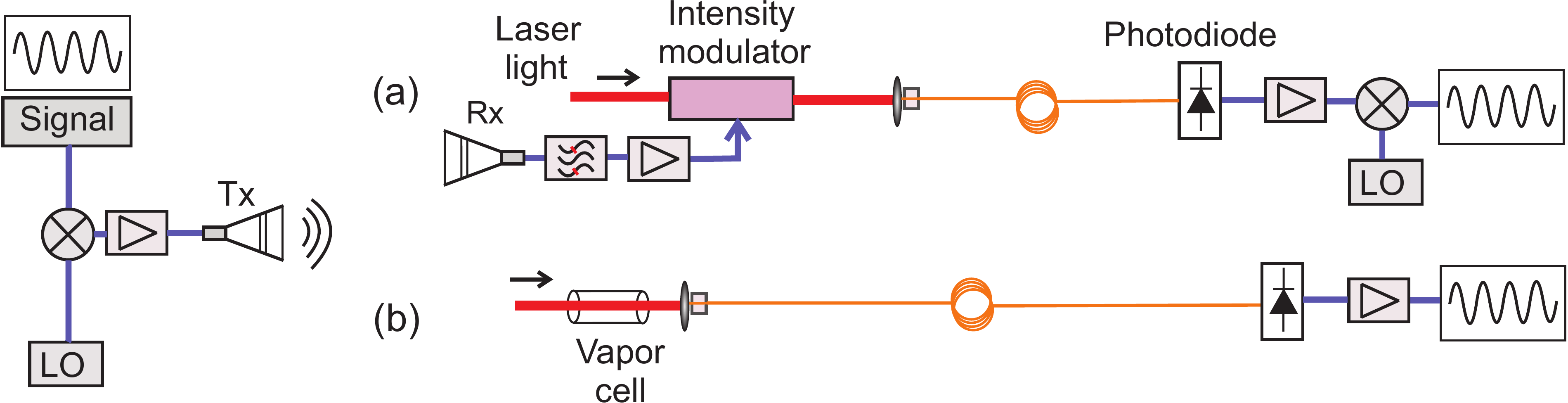}
\caption{\label{fig:wide} (a) A signal is mixed with a local oscillator (LO) and coupled to free-space via a transmitter (Tx) antenna. (b) In a conventional RoF architecture, the mixed signal from the receiver antenna (Rx) is amplified and used to intensity-modulate a laser beam. After travelling through optical fiber the light is received on a photodetector. The photodetector signal is demodulated at the LO frequency to retrieve the baseband signal. (c) The RoF realized in this work. Atoms in a vapor cell encode the rf signal into a light beam. Photodetection following an optical fiber, directly retrieves the baseband signal.}
\end{figure*}

Rydberg atoms have become a prominent area of research in atomic and optical physics in the recent years \cite{Saffman2008,Firstenberg2015}. Those highly excited states of atoms with a principal quantum number $n > 30$ possess very high ac polarizabilities and therefore are extremely sensitive to oscillating electric fields \cite{Gallagher1994}. Electromagnetically-induced transparency (EIT) in three-level atomic systems - where quantum interference renders an opaque atomic medium transparent when a resonance condition is met - has been widely studied in the last two decades \cite{Harris1990,Fleischhauer2005}. EIT as means of coherent optical detection of  Rydberg states demonstrated in the pioneering work of \cite{Adams2007} paved the way towards breakthrough experiments on giant Kerr nonlinearities with Rydberg atomic media \cite{Mohapatra2008}, quantum nonlinear optics at the single-photon level \cite{Firstenberg2015,Peyronel2012}, optical electrometry of microwave fields \cite{Shaffer2012,Holloway2014} and coherent microwave-to-optical conversion \cite{Han2018}.


In this Letter, we demonstrate a sensitive radio-over-fiber architecture using an optical receiver antenna for microwaves using Rydberg states of atoms in a vapor cell. By exposing atoms to suitable microwave fields and establishing an electromagnetically induced transparency (EIT) resonance condition, we transferred analog rf signals to the optical domain. A fiber-optical channel links the light carrying the signal to a photodetector. Here the signal is retrieved to the baseband rf domain without requiring any demodulation stage. We demonstrated the functionality of the optical antenna over a number of local oscillator (LO) frequencies, namely 12.7 GHz, 14.2 GHz, 17 GHz, and 19.3 GHz, for an analog modulation bandwidth up to $\sim$1.2~MHz. 

\begin{figure}
\includegraphics[width=\columnwidth]{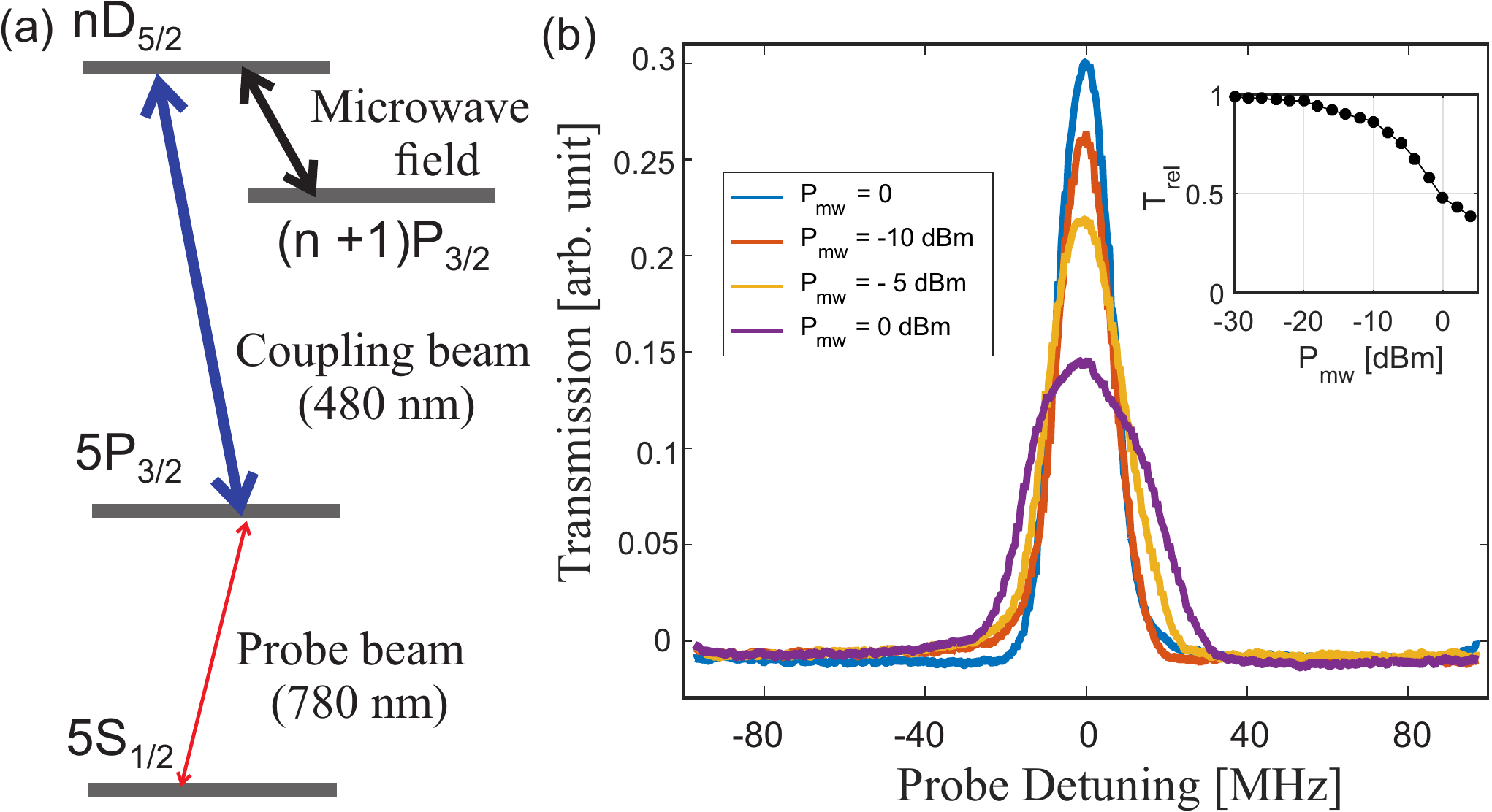}
\caption{\label{fig:wide} (a) The atomic level structure of Rb used in this work (see text). (b) Effect of the carrier microwave at 19.3 GHz on the transmission of a weak probe beam across the resonance at various levels of microwave power P$_{\mathrm{mw}}$ fed into the antenna. The inset shows the relative transmission as a function P$_{\mathrm{mw}}$. }
\end{figure}


The heart of our optical antenna is an atomic vapor cell containing rubidium atoms. The atomic structure relevant for our experiments is shown in figure 2a. An intense laser beam (the coupling beam), is frequency-tuned close to the resonance between the first excited state (5P$_{3/2}$) and a particular Rydberg state ($n$D$_{5/2}$, $n$~=~45-57 in this work). Atoms in their ground state (5S$_{1/2}$) are interrogated by a weak probe beam tuned across the 5S$_{1/2} \leftrightarrow$ 5P$_{3/2}$ transition. In the presence of the strong coupling beam, a transparency window opens at the probe beam resonance leading to an enhanced transmission of the probe beam through the vapor cell. If a microwave field couples the Rydberg state $n$D$_{5/2}$ to another nearby Rydberg state, the EIT resonance is split into two peaks (Autler-Townes splitting \cite{Tannoudji}). In this work, we couple to the state $(n+1)$P$_{3/2}$,  and the amount of splitting varies linearly with the microwave field.  Due to extremely high ac polarizability of Rydberg states, even weak microwave fields can lead to a substantial Autler-Townes splitting, thereby modifying the EIT resonance window. Amplitude modulation of the microwave field hence causes a modulation of the transmitted probe light opening up the possibility of encoding the amplitude and phase of a signal RF field in the optical domain. In effect, an optical antenna is realized where a laser beam is amplitude-modulated by an rf signal. The laser light relays the encoded information to a photodetector that generates a photocurrent proportional to the optical intensity. A transimpedance amplifier converts photocurrent to a photovoltage that contains the original signal. Using different pairs of Rydberg states, we were able to encode the rf signal into different carrier microwave fields: 48D$_{5/2} \leftrightarrow$ 49P$_{3/2}$~(19.3\,GHz), 50D$_{5/2} \leftrightarrow$ 51P$_{3/2}$~(17\,GHz), 53D$_{5/2} \leftrightarrow$ 54P$_{3/2}$~(14.2\,GHz), and 55D$_{5/2} \leftrightarrow$ 56P$_{3/2}$~(12.7\,GHz) without needing any physical reconfiguration of the optical antenna.

\begin{figure*}
\includegraphics[width=177mm]{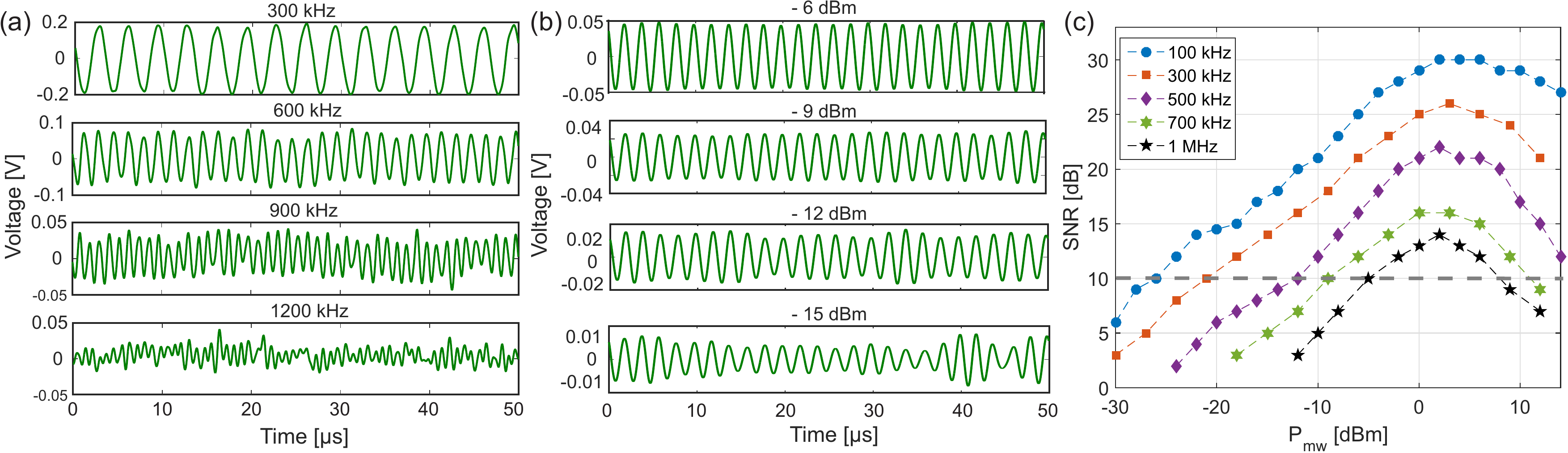}
\caption{\label{fig:wide} (a) Retrieved sinusoidal rf signals for various modulation frequencies for a 19.3~GHz carrier with power P$_{\mathrm{mw}}$ = -1~dBm. (b) Retrieved rf signal for a modulation frequency of 500~kHz for various power levels of the 19.3~GHz carrier. (c) Signal-to-noise ratio (see text) as a function of the carrier power for a range of modulation frequencies. }
\end{figure*}

In our experimental setup, the coupling laser light at 480 nm and the probe light at 780 nm are derived from external-cavity diode lasers (ECDL Toptica DL-pro). About 25 mW of blue light is focussed to a 90~$\mu$m spot size radius in the vapor cell. Counter-propagating to the coupling beam we have a probe beam focussed into an 80~$\mu$m spot size radius and having typically about 4~$\mu$W of power. The beams are combined and separated using dichoric mirrors and are both vertically polarized. The vapor cell is heated up to a temperature of $60^{\mathrm{o}}\mathrm{C}$ to increase vapor pressure. By coarse tuning of the blue laser, a range of Rydberg states $n$D$_{5/2,3/2}$ can be accessed. Microwave fields were sourced from a signal generator (Agilent EE825D) and were coupled to the free space using a home-built helical end-fire antenna with an estimated gain of $\sim$18 dBi at $\sim$15 GHz. This antenna produces circularly-polarized microwave fields is used because of its simplicity of construction. The emitter antenna was placed 0.5~meter away from the vapor cell and a microwave power of up to 12 dBm can be provided to the antenna. Assuming that the atoms are located along the axial direction of the antenna and that no reflection of the microwave field from nearby objects is present, a power-feed of -15\,dBm in the antenna corresponds to a field intensity of $\sim$ 1 $\mu$W/cm$^2$ at the location of the atoms.  After passage through the vapor cell, the weak probe light is coupled to a multimode optical fiber with an efficiency of 85$\%$ and is detected on a fast (100 MHz bandwidth), fiber-coupled ac photodetector (Thorlabs PDB415A). A high-pass filter with a 100 kHz cutoff point eliminates a weak 75\,kHz modulation present in the probe beam originating from a lock-in amplifier used for frequency-locking the laser. For this work, shot-noise limited performance of the detector was not achieved. Figure 2b shows the background-subtracted transmission of the probe beam across the 5S$_{1/2} \leftrightarrow$ 5P$_{3/2}$ transition in the presence of the coupling beam tuned to the 5P$_{3/2}\leftrightarrow$ 48D$_{5/2}$  transition. In the presence of a microwave field at frequency 12.7\,GHz, the transmission drops. The Autler-Townes spectrum is not resolved at low microwave powers in our case due to the use of circularly polarized microwave fields, which causes the vector nature of the ac polarizability to play a role and gives rise to a central peak that decreases with increasing microwave power \cite{Sedl2013, note2}. The inset shows the relative transmission of the probe at resonance as a function of the microwave power.

The dependence of the light transmission on the microwave field strength allows one to modulate the light transmission by modulating the carrier microwave field using an rf signal field. To achieve this, a test sinusoidal rf signal was generated by an arbitrary waveform generator (Agilent 33120A) and was used to provide direct analog amplitude modulation to the microwave carrier. The amplitude and the phase of the signal field are thus encoded in the microwave field that the atoms in the vapor cell are exposed to. The applied rf field is adjusted such that its full amplitude corresponds to a modulation depth of 100\% for the microwave field. Figure 3a shows single-shot traces of the photovoltage produced by the transmitted probe light for a range of modulation frequencies for a carrier microwave field at 19.3 GHz amplitude-modulated by sinusoidal signal rf fields. Evidently, high signal fidelity can be maintained with modulation frequencies $\sim$\,MHz. Figure 3b shows the photovoltages for a modulation frequency of 300\,kHz for a range of carrier microwave powers. In order to quantify the signal-to-noise (SNR) in the photovoltages produced by the transmitted light, we measured them on a spectrum analyser with a resolution bandwidth of 10\,kHz. We defined the SNR as the ratio of the power spectral density at the modulation frequency to the average power spectral density of the electronic noise floor over a 2 MHz band. The gain flatness of the photodetector is better than 1.5 dB over this band. Figure 3c shows the SNR as a function of the carrier power P$_{\mathrm{mw}}$ for a range of modulation frequencies. The SNR grows approximately linearly with the carrier power up to about 3 dBm beyond which it drops. Defining SNR = 10 as the cutoff point for high-fidelity signal transfer (the grey horizontal line in Figure 3c), we observe that it can be achieved for a signal bandwidth of 1 MHz over a range of P$_{\mathrm{mw}}$. The noise in our setup is dominated by electronic noise of the transimpedance stage of the photodetector and the laser frequency noise. We note that Rydberg states are sensitive to low-frequency ($<$ 1 kilohertz) electric fields which cause a dc Stark shift of these states. Any such slow-varying fields, however, do not affect our detection since the feedback loop used for frequency-locking the coupling laser automatically tracks such variations. Background electric field amplitude at a frequency close to the signal bandwidth is typically extremely small ($<$ 10 nV/cm) and can be further reduced by shielding low-frequency electric fields [cite].


Figure 4 shows the SNR as a function of the modulation frequency for three different microwave carrier frequencies, corresponding to different pairs of Rydberg states coupled by the microwave field. The coupling beam power is kept constant at 25\,mW, as is the carrier microwave power (-1\,dBm). The data suggest that the maximum signal bandwidth for our implementation of RoF is limited to $\sim 1.1$~MHz. The cutoff modulation frequency (grey horizontal line) appears to be weakly dependent on the carrier microwave frequency being used. The inset shows the cutoff frequency as a function of the coupling beam power for a 14.2~GHz carrier field with a power of -1~dBm. Evidently, the maximum available signal bandwidth in our regime of experimental parameters increases nearly linearly with the coupling beam power, suggesting that we are currently limited by the available laser power.


\begin{figure}[tb!]
\centering
\includegraphics[width=1\columnwidth]{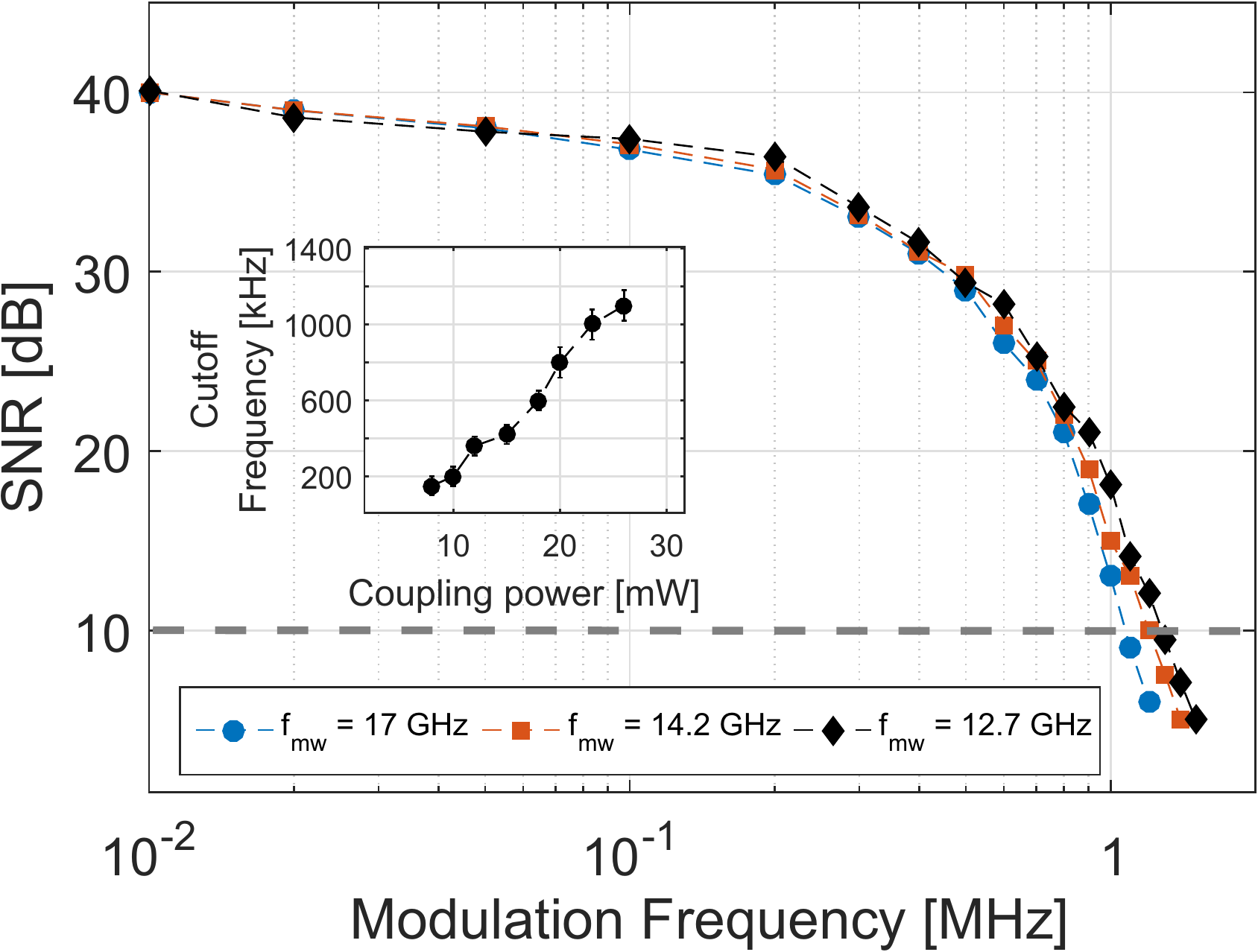}
\caption{\label{fig:wide} Signal-to-noise as a function of modulation frequency for a range of microwave carrier frequencies. The inset shows the cutoff modulation frequency (defined in the text) for $f_{mw}$ = 14.2~GHz as a function of coupling beam power.}
\end{figure}

In order to estimate the fundamental limit on the achievable signal bandwidth for RoF, we note that the enhanced transmission near EIT arises from the population of a dark state which is a dressed eigenstate of an atom in the presence of the strong coupling laser field. The lifetime of atoms in the dark state determines how fast those states can be depopulated once they are occupied by the atoms, which in turn sets a limit on RoF bandwidth. The spectroscopic width of the dark state (the EIT linewidth), by the uncertainty principle, is the inverse of its lifetime. In real experiments the EIT width is governed by a number of parameters, including the coupling laser and microwave intensity, probe beam size, the natural linewidth of the probe transition, thermal dephasing owing to atoms moving in and out of the EIT zone of the vapor cell and collisional dephasing. A full model describing the effects of these parameters on the achievable RoF sensitivity and bandwidth is beyond the scope of this Letter. Qualitatively, in the limit of no thermal or collisional dephasing and a weak probe field, the maximum achievable EIT linewidth while ensuring sufficiently enhanced transmission is of the order $\Gamma$, where $\Gamma$ is the natural linewidth of the probe transition \cite{Gea1995}. The minimum decay time of the dark states is thus in the order of $\sim 1/\Gamma$, which for our probe transition is about 26 ns. In other atomic species, the first excited state lifetime can be much smaller (in atomic strontium, e.g., this is less than 5~ns). This limit can be surmounted by exploiting collective effects in an ensemble of atoms. Dicke superradiance \cite{Dicke1954,Haroche1982}, for instance, can be used to reduce the excited state decay time to values much smaller than $1/\Gamma$. This requires dense atomic samples with $\rho\/k^3  \sim 1 $, where $\rho$ is the atomic density and $k = 2\pi/\lambda$, where $\lambda$ is the transition wavelength. Optical line broadening mechanisms in dense, cold ensembles due to dipole-dipole interactions is a topic of significant recent interest \cite{Browaeys, JunYe, Bettles2016}. Even in dilute atomic ensembles, cooperative effects can lead to superradiance-like decay in the forward direction. Coherent emission of pulse trains of lengths a hundred times shorter than $1/\Gamma$ has recently been demonstrated \cite{Wilkowski1}. It may be possible, in principle, to use our optical atomic antenna for RoF signal bandwidths well in the giga-hertz domain.

In conclusion, we have demonstrated a proof-of-concept radio-over-fiber signal delivery system using an optical antenna based on atoms in a vapor cell. A signal bandwidth exceeding 1\,MHz is achieved and is limited by technical noise. Higher bandwidths in our current implementation can be achieved through a combination of shot-noise limited photodetection and higher coupling laser power. In addition, SNR and bandwidth can be further enhanced by interferometric techniques e.g., frequency-modulation spectroscopy \cite{bjorklund} and measuring optical response in the dispersive domain. We point out that in many applications of RoF, signal bandwidth is not the sole figure of merit. Electrical isolation is a prime concern for geophysical applications in order to avoid fatal damage caused by e.g., thunderbolts and solar storms which our RoF implementation is inherently immune to. Complete electrical isolation also eliminates corruption of the signal by unwanted electrical contacts in the coax and its exposure to external sources of radiation. The vapor cell and fiber assembly (which can be further miniaturized) is considerably more lightweight than conventional RoF designs and can be advantageous in certain applications such as avionics \cite{Novak2016}. The nonmetallic nature of our optical antenna and the fiber-optic signal transport system can further be beneficial for electrophysiological signal detection in medical environments. Indeed, in such applications, a signal bandwidth of DC - 1\,MHz, as  in this demonstration, can be a widely usable range.



While completing this Letter, we became aware of a similar work on digital communication with Rydberg atoms \cite{meyer2018}. Subsequent to \cite{meyer2018} and the submission of our Letter, a manuscript describing a Rydberg atom-based quantum receiver appeared \cite{raithel2018}.

\begin{acknowledgments}
We acknowledge Otago Innovation Limited, the Marsden Fund of New Zealand (contract UOO1729), and a University of Otago Research Grant (UORG) for financial support. We thank Craig Chisholm for calculating the properties of Rydberg atoms.
\end{acknowledgments}



\begin{thebibliography}{99}
\bibitem{note}
The demodulation stage can alternatively be before the rf-to-optical link. Either of the configurations is prone to the same complexity and reconfigurability issues.
\bibitem{note2}
A circularly-polarized field makes certain Zeeman hyperfine electric dipole transitions allowed for the Rydberg states. For a linearly-polarized coupling and probe beams (as in our case), this causes several hyperfine states of the ground and the first excited electronic states to play a role in the overall transmission of light. This leads to a three-peak, rather than a typical two-peak Autler-Townes EIT spectrum.
\end{thebibliography}

\end{document}